\documentclass[a4paper,12pt]{article}
\usepackage{soul}
\usepackage{amsthm}
\newtheorem{example}{Example}
\newtheorem*{Def}{Definition}

\usepackage[dvipsnames]{xcolor}
\usepackage{mathrsfs}

\usepackage{subfig}

\usepackage{setspace}
\usepackage{hyperref}
\usepackage{amsmath}
\usepackage{amssymb}
\usepackage{multirow}
\usepackage{booktabs}
\usepackage{graphicx}
\usepackage{multirow}
\usepackage{natbib}
\usepackage{comment}

\usepackage{graphicx}
\usepackage {tikz}
\usetikzlibrary {shapes ,arrows , positioning }

\usepackage[margin=1in]{geometry}
\usepackage{authblk}

\title{\textbf{On integral priors for multiple comparison in Bayesian model selection\thanks{Accepted for publication in \emph{International Statistical Review}
			(published by Wiley). The final version of record is available at
			https://doi.org/10.1111/insr.70028}}}

\author[1,2,3]{Diego Salmerón}
\author[4]{Juan Antonio Cano}
\author[5,6]{Christian P. Robert}

\affil[1]{\small{\textit{Health and Social Sciences Department, University of Murcia, 30120, Spain}}}
\affil[2]{\small{\textit{CIBER de Epidemiología y Salud Pública (CIBERESP), 28029, Madrid, Spain}}}
\affil[3]{\small{\textit{Biomedical Research Institute of Murcia Pascual Parrilla–IMIB, 30120 Murcia, Spain}}}
\affil[4]{\small{\textit{Department of Statistics and Operational Research, University of Murcia, 30100, Spain}}}
\affil[5]{\small{\textit{Université Paris Dauphine PSL, 75775 Paris cedex 16, France}}}
\affil[6]{\small{\textit{Department of Statistics, University of Warwick, Coventry CV4 7AL, United Kingdom}}}

\usepackage{caption}
\DeclareCaptionLabelFormat{boldfig}{\textbf{#1~#2}} % Solo cambia Figure
\begin{document}
%	\author{Diego Salmer\'on$^*$, Juan Antonio Cano$^{*}$\footnote{$^{*}$ Juan Antonio Cano (1956-2018) was Professor in the Department of Statistics and Operations Research at the University of Murcia. Dr. Cano was a dear mentor and friend who contributed substantially to the theory of integral priors and was instrumental to the developments of this article.}\\and Christian P. Robert$^{\dagger\ddagger}$\\ 
%		\\
%		\small{$^{*}$Universidad de Murcia, Spain}, \small{$^{\dagger}$Universit\'e Paris Dauphine PSL, France}\\
%		\small{$^{\ddagger}$University of Warwick, United Kingdom}
%	}
	\date{}
	\maketitle
\doublespacing
\begin{abstract}
Noninformative priors constructed for estimation purposes are usually not appropriate for model selection and testing. The methodology of integral priors was developed to get prior distributions for Bayesian model selection when comparing two models, modifying initial improper reference priors. We propose a generalization of this methodology to more than two models. Our approach adds an artificial copy of each model under comparison by compactifying the parametric space and creating an ergodic Markov chain across all models that returns the integral priors as marginals of the stationary distribution. Besides the guarantee of their existence and the lack of paradoxes attached to estimation reference priors, an additional advantage of this methodology is that the simulation of this Markov chain is straightforward as it only requires simulations of imaginary training samples for all models and from the corresponding posterior distributions. %This renders its implementation automatic and generic, both in the nested and in the non-nested cases. 
We present some examples, including situations where other methodologies need specific adjustments or do not produce a satisfactory answer.
\newline

\noindent
\textbf{Keywords}: Integral priors; Bayesian model selection; Objective Bayes factor; Markov chains.
\end{abstract}

\section{Introduction}

Noninformative (or objective) priors are essential tools for Bayesian analysis, even though the denominations of `noninformative' and `objective' are often criticised as being inappropriate \citep{bad}. The main motivation is in producing a reference prior distribution that impacts as little as possible the resulting posterior distribution \citep{BergerBernardoSun2024}. Further motivations range from closing the inference space as in complete class results \citep{Robert2001} to achieving better frequentist properties such as asymptotic coverage \citep{rousseau:07}.
The main issue with such priors is that there exists an extensive literature on the possible choices and theories of noninformative priors. Following their formalisation by Harold Jeffreys \citep{Jeffreys1939}, several propositions emerged towards a theory of such priors, with a varying degree of generality as reported in \cite{kass:wasserman:1995}, \cite{Robert2001}, \cite{riss}, and \cite{BergerBernardoSun2024}. The most elaborate approach to this day is the concept of reference priors developed by the latter (see also \cite{Bernardo1979} and \cite{BergerBernardo1992}). 

These different methods most usually yield an improper prior measure, that is, a measure with infinite mass, which is thus determined up to a positive multiplicative constant since it cannot be normalised into a probability distribution. Most often this feature is not a serious issue for estimation purposes because the constant cancels in the posterior distribution. Unfortunately, in model selection problems such as hypothesis testing where two (or more) models compete for consideration, $M_i:$ $\{f_i(x\mid\theta_i)$, $\theta_i\in\Theta_i\}$, $i=1,2$, resorting to improper priors $\pi_i^N(\theta_i)=c_i h_i(\theta_i)$, $i=1,2$, implies that the optimal Bayesian procedure, namely the Bayes factor \citep{Jeffreys1939}
\[
B_{21}^N(x)=\frac{c_2\int f_2(x\mid\theta_2)h_2(\theta_2)\mathrm{d}\theta_2}{c_1\int f_1(x\mid\theta_1)h_1(\theta_1)\mathrm{d}\theta_1},
\]
depends on the arbitrary ratio $c_2/c_1$. Several general approaches have been proposed for prior construction in Bayesian model selection, among which \cite{BayarriGarciaDonato2008}, \cite{Bayarri2012}, and \cite{Fouskakis2015}. A review of reference priors for model selection can be found in \cite{Consonni2018} as well as in \cite{BergerBernardoSun2024}.

One of these attempts towards alleviating this issue is the methodology of intrinsic priors \citep{BergerPericchi1996} that exploit a (minimal) fraction of the data to update the improper priors into proper posteriors. In short, intrinsic priors are defined as the solution $\{\pi_1^I(\theta_1),\pi_2^I(\theta_2)\}$ of a system of two functional equations. Under certain conditions, and assuming that $M_1$ is nested in $M_2$, the system of functional equations reduces to a single equation with two functional unknowns, in such a way that once an arbitrary choice is made for $\pi_1^I(\theta_1)$, the dual prior $\pi_2^I(\theta_2)$ is automatically determined. The solution $\{\pi_1^I(\theta_1),\pi_2^I(\theta_2)\}$ proposed by \cite{Moreno1998} consists of choosing $\pi_1^I(\theta_1)=\pi_1^N(\theta_1)$ and hence
\[
\pi_2^I(\theta_2)=\pi_2^N(\theta_2)\mathbb E_{f_2(z\mid\theta_2)}(B_{12}^N(Z))=\int \pi_2^N(\theta_2\mid z)m_1^N(z)\mathrm{d}z,
\]
where $z$ denotes an imaginary training sample, $B_{12}^N(z)=m_1^N(z)/m_2^N(z)$, and $m_i^N(z)=\int f_i(z\mid\theta_i)\pi_i^N(\theta_i)\mathrm{d}\theta_i$, $i=1,2$. See \cite{PerezBerger2002} for more details. However, intrinsic Bayes factors are often not suitable in the non-nested case, see \cite{BergerMortera1999} and \cite{Moreno2005}. Another proposal consists of using fractional priors, which are again defined as the solution to a system of functional equations (see \cite{ohagan:1997} and  \cite{Moreno1997}). However, this methodology has received less attention and also faces difficulties \citep{Robert2001}.

\cite{PerezBerger2002} have proposed a novel approach under the name of expected posterior priors distributions. This method can be applied in the general scenario where we have $q\geq 2$ models under consideration. The expected posterior priors are defined as
\[
\pi_i^{*}(\theta_i)=\int\pi_i^N(\theta_i\mid z)m^{*}(z)\mathrm{d}z,\,\,\,i=1,\dots,q,
\] 
where $m^{*}(z)$ is an arbitrary predictive density, which can be improper, for the imaginary training sample $z$. When $M_1$, say, is nested in the other models, \cite{PerezBerger2002} propose  $m^{*}(z)=m_1^N(z)$, and therefore this method applied to two nested-encompassing models produces the intrinsic priors introduced by \cite{Moreno1998}.

\subsection{Integral priors}
Integral priors have been introduced in \cite{Cano2008} for model selection when two models, $M_i:$ $\{f_i(x\mid\theta_i)$, $\theta_i\in\Theta_i\}$, $i=1,2$, are under consideration. These prior distributions, $\pi_1(\theta_1)$ and $\pi_2(\theta_2)$, are defined as the solution of a system of integral equations,  and can be considered as a generalization of the expected posterior priors introduced in \cite{PerezBerger2002}. Concretely, integral priors $\{\pi_1(\theta_1),\pi_2(\theta_2)\}$
satisfy the integral equations
\[
\pi_1(\theta_1)=\int\pi_1^N(\theta_1\mid z_1)m_2(z_1)\mathrm{d}z_1,
\]
\[
\pi_2(\theta_2)=\int\pi_2^N(\theta_2\mid z_2)m_1(z_2)\mathrm{d}z_2,
\]
where $\pi_i^N(\theta_i)$ is a prior distribution for estimation (typically a noninformative prior), $\pi_i^N(\theta_i\mid z)\propto
f_i(z\mid\theta_i)\pi_i^N(\theta_i)$, $m_i(z)=\int
f_i(z\mid\theta_i)\pi_i(\theta_i)\mathrm{d}\theta_i$, $i=1,2$, 
and $z_1$
and $z_2$ are (minimal) imaginary training samples for $\theta_1$ and $\theta_2$, respectively. When $\theta_i$ or $z_i$ are discrete, the corresponding integrals are replaced by sums  with no loss of generality. These integral equations balance each model with respect to the other one since the prior $\pi_i(\theta_i)$
is derived from the marginal $m_j(z_i)$, and therefore from $\pi_j(\theta_j)$, $j\neq i$, as an unknown generalized expected posterior prior.

Integral priors can be used to compare both nested and non-nested models without the need to make this distinction, and the application of this methodology does not need to specify a predictive distribution as with the expected posterior priors
approach. 

Some examples where the exact form of integral priors is known are studied in \cite{Cano2008}. The simplest one is the point null hypothesis testing $H_0:\theta=\theta^*$ \textit{versus} $H_1:\theta\neq\theta^*$, for which the integral priors are $\pi_1(\theta_1)=\delta_{\theta^*}(\theta_1)$ and $\pi_2(\theta_2)=\int\pi_2^N(\theta_2\mid z)f_1(z\mid\theta^*)\mathrm{d}z$, respectively. A further example is the comparison of two location models, where the priors $\pi_i(\theta_i)=\pi_i^N(\theta_i)=1$, $i=1,2$, are integral priors. This extends to the comparison of two scale and two location-scale models, the initial default priors, $\pi_i(\sigma_i)$ $=\pi_i^N(\sigma_i)=1/\sigma_i$, $i=1,2$, for the former, and $\pi_i(\mu_i,\sigma_i)=\pi_i^N(\mu_i,\sigma_i)$ $=1/\sigma_i$, $i=1,2$, for the latter, being integral priors. Further examples have been provided in \cite{Cano2008}, while significant differences between intrinsic and integral priors have been discussed in \cite{Cano2018}.

In general closed-form solutions to the integral equations are unavailable. To deal with this drawback we established an equivalence between the solution of the integral equations and the convergence of two Markov chains. 
%\cite{Cano2008} showed that $\pi_1(\theta_1)$ and $\pi_2(\theta_2)$ are integral priors if and only if $\pi_1(\theta_1)$ and $\pi_2(\theta_2)$ are the invariant $\sigma$-finite measures of two Markov chains, $(\theta_1^t)$ and $(\theta_2^t)$, respectively.
Specifically, \cite{Cano2008} considered Markov chains $(\theta_1^t)$ and $(\theta_2^t)$ whose $\sigma$-finite invariant \textit{distributions} are the integral priors.

 The transition $\theta_1^{\prime}\rightarrow\theta_1$ of the Markov chain $(\theta_1^t)$ is the marginal of the following conditional steps:
\begin{enumerate}
	\item Simulate a training sample $z_2 \sim f_1(z_2\mid\theta_1^{\prime})$
	\item Simulate $\theta_2\sim\pi_2^N(\theta_2\mid z_2)$
	\item Simulate a training sample $z_1 \sim f_2(z_1\mid\theta_2)$
	\item Simulate $\theta_1 \sim \pi_1^N(\theta_1\mid z_1)$,
\end{enumerate}
and the transition $\theta_2^{\prime}\rightarrow\theta_2$ of the Markov chain $(\theta_2^t)$ is obtained in a symmetric manner, with both chains being simultaneously recurrent or transient. Therefore, the study of integral priors can be performed considering the stochastic stability of these Markov chains, as in \cite{Cano2007a}, \cite{Cano2007b}, \cite{Cano2013}, \cite{Salmeron2015}, \cite{Cano2016}, and \cite{Cano2018}. 

In some situations the Markov chains attached to integral priors are positive recurrent, and even sometimes positive Harris recurrent, in which case the integral priors are proper prior distributions. This implies that the Bayes factor $B_{21}(x)$ is defined without ambiguity \citep{Robert2001} and can be estimated by the Monte Carlo estimator
\begin{equation}
\hat{B}_{21}(x)=\frac{\sum_{t=1}^T f_2(x\mid\theta_2^t)}{\sum_{t=1}^T f_1(x\mid\theta_1^t)}\label{MCestimator}
\end{equation}
associated with the above Markov chains.

In the general case \cite{Cano2013} have proposed to consider constrained imaginary training samples $z$ when the simulation of the chains is performed, which guarantees the existence and the uniqueness of invariant probability measures. This approach has been implemented successfully in several situations, see \cite{Cano2013}, and \cite{Cano2018}. However, the simulation of constrained imaginary training samples can involve high computing time, specially when the accept-reject method is implemented. In this article we propose a new approach that circumvents the need to simulate constrained imaginary training samples and that enables the methodology to be extended to comparing more than two models at once.

The plan of this paper is as follows. In Section 2 we generalize the definition of integral priors for more than two models. In Section 3 we propose to add copies of the models under consideration by introducing artificial compact parametric spaces, and we show that integral priors are proper priors. Section 4 is dedicated to show how the methodology applies to the problem of testing if the mean of a normal population is negative, zero, or positive. This multiple non-nested hypothesis testing problem has previously been studied in \cite{BergerMortera1999} assuming known variance when using both intrinsic and fractional Bayes factor. However, as pointed out by \cite{BergerMortera1999}, these approaches do not correspond to genuine Bayes factors attached to specific prior distributions, and there is no a general methodology that, when applied to this model selection problem, produces priors with a satisfactory answer. In Section 5 we apply our methodology to the one way ANOVA model. Section 6 is dedicated to the problem of testing the Poisson versus the negative binomial using integral priors; a model selection problem where the definition of intrinsic priors presents a certain degree of arbitrariness when the initial default prior for estimation is the natural choice, that is, the Jeffreys priors. In Section 7 we show how the simulation from the integral posterior can be performed. Finally, in Section 8 we present some relevant conclusions and outline incoming research.

\section{The Markov chains associated with integral priors}
%When considering only two models, the integral priors are the invariant $\sigma$-finite measures of two Markov chains, $(\theta_1^t)$ and $(\theta_2^t)$, respectively. 
In this section we extend the definition of the integral priors when comparing more than two models, through a generalization of the above Markov chains.

As a starter, it proves convenient to represent the simulation of the dual chains $(\theta_1^t)$ and $(\theta_2^t)$ as shown in Figure  \ref{basic_definition}, where each single step $\theta_i\rightarrow\theta_j$, $i\neq j$, is carried out by first simulating $z_j\sim f_i(z_j\mid\theta_i)$, which we call an imaginary training sample of sufficient size to make $\pi_j^N(\theta_j\mid z_j)$ proper, and second simulating $\theta_j\sim\pi_j^N(\theta_j\mid z_j)$, that is, simulating from the marginal transition
\begin{equation}
	\pi_{ij}(\theta_i,\theta_j)=\int f_i(z_j\mid\theta_i)\pi_j^N(\theta_j\mid z_j)\mathrm{d}z_j\label{singleTransition}.
\end{equation}
More specifically, given an initial point $\theta_1^1\in\Theta_1$, we carry out the transitions $\theta_1^1\rightarrow\theta_2^1\rightarrow\theta_1^2\rightarrow\theta_2^2\,\cdots$ obtaining two marginal (interleaved) Markov chains, $(\theta_1^t)$ and $(\theta_2^t)$.
\begin{figure}[b]
	\centering
	\begin{tikzpicture}[->,>= stealth',shorten >=2pt,line width =0.5 pt,node distance =2 cm,style ={minimum size=5mm},bend left=12]
		\node [] (2) at (0, 0) {$\theta_2^t$};
		\node [] (1) at (0, 2) {$\theta_1^t$};
		\node [] (2p) at (2, 0) {$\theta_2^{t+1}$};
		\node [] (1p) at (2, 2) {$\theta_1^{t+1}$};
		\draw[->] (1) -- (2);
		\draw[->] (2) -- (1p);
		\draw[->] (1p) -- (2p);
	\end{tikzpicture}
	\caption {\textit{Transition of the Markov chains for two models under comparison}.}
	\label{basic_definition}
\end{figure}
When starting with an initial point $\theta_2^1\in\Theta_2$, the process maintains the same \textit{stochastic behaviour}. 

However, with more than two models under comparison, there exist several possibilities to combine simulations from $\pi_{ij}(\theta_i,\theta_j)$ in a sequence of simulation steps, towards obtaining Markov chains on each parametric space. Consider the case with three models. Figure \ref{Possibility} shows two scenarios to combine single steps. It is worth noting the similarity with Gibbs sampling. However, contrary to what happens with Gibbs sampling where the chosen order in which we carry out the simulation does not impact the stationary distribution (albeit it does impact the convergence rate), in the context we are dealing with,  order matters and each scenario can produce a Markov chain on $\Theta_i$ with a different stationary distribution, $i=1,2,3$. Suppose that $M_1$ corresponds to the point null
hypothesis $\theta_1=\theta_1^*$. Therein the stationary distribution $\pi_2(\theta_2)$ is
$\pi_{12}(\theta_1^*,\theta_2)$
under scenario 1, and $\int \pi_{32}(\theta_3,\theta_2)\pi_{13}(\theta_1^*,\theta_3)\mathrm{d}\theta_3$
under scenario 2, which obviously differ. Hence the specific way in which the single steps are combined can produce different answers, with the consequent problem of defining integral priors for more than two models.
\begin{figure}[t]
	\centering
	\begin{tikzpicture}[->,>= stealth',shorten >=2pt,line width =0.5 pt,node distance =2 cm,style ={minimum size=5mm},bend left=12]
		\node [] (0) at (1,5) {\,\,\,\,\,\,\,\,\,\,\,\, Scenario 1};
		\node [] (1) at (0, 4) {$\theta_{1}^t$};
		\node [] (2) at (0, 2) {$\theta_{2}^t$};
		\node [] (3) at (0, 0) {$\theta_{3}^t$};
		\node [] (1p) at (3, 4) {$\theta_{1}^{t+1}$};
		\node [] (2p) at (3, 2) {$\theta_{2}^{t+1}$};
		\node [] (3p) at (3, 0) {$\theta_{3}^{t+1}$};
		\draw[->] (1) -- (2);
		\draw[->] (2) -- (3);
		\draw[->] (3) -- (1p);
		\draw[->] (1p) -- (2p);
		\draw[->] (2p) -- (3p);
		\node [] (0) at (7,5) {\,\,\,\,\,\,\,\,\,\,\,\, Scenario 2};
		\node [] (1) at (6, 4) {$\theta_{1}^t$};
		\node [] (2) at (6, 2) {$\theta_{3}^t$};
		\node [] (3) at (6, 0) {$\theta_{2}^t$};
		\node [] (1p) at (9, 4) {$\theta_{1}^{t+1}$};
		\node [] (2p) at (9, 2) {$\theta_{3}^{t+1}$};
		\node [] (3p) at (9, 0) {$\theta_{2}^{t+1}$};
		\draw[->] (1) -- (2);
		\draw[->] (2) -- (3);
		\draw[->] (3) -- (1p);
		\draw[->] (1p) -- (2p);
		\draw[->] (2p) -- (3p);
	\end{tikzpicture}
	\caption {\textit{Two competing scenarios to combine conditional updating steps when comparing three models}.}
	\label{Possibility}
\end{figure}
A straightforward proposal that makes this issue vanish is to randomly choose the ordering in which the steps are performed: if we have carried out the steps  $\theta_{i_1}^t\rightarrow\theta_{i_2}^t\rightarrow\theta_{i_3}^t$ for a given sequence $(i_1,i_2,i_3)$, then we select at random a sequence $(j_1, j_2, j_3)$ with the condition that $j_1\neq i_3$, and then we perform the steps $\theta_{i_3}^t\rightarrow\theta_{j_1}^{t+1}\rightarrow\theta_{j_2}^{t+1}\rightarrow\theta_{j_3}^{t+1}$, as shown in Figure \ref{new_definition}.
\begin{figure}[!b]
	\centering
	\begin{tikzpicture}[->,>= stealth',shorten >=2pt,line width =0.5 pt,node distance =2 cm,style ={minimum size=5mm},bend left=12]
		\node [] (1) at (0, 4) {$\theta_{i_1}^t$};
		\node [] (2) at (0, 2) {$\theta_{i_2}^t$};
		\node [] (3) at (0, 0) {$\theta_{i_3}^t$};
		\node [] (1p) at (3, 4) {$\theta_{j_1}^{t+1}$};
		\node [] (2p) at (3, 2) {$\theta_{j_2}^{t+1}$};
		\node [] (3p) at (3, 0) {$\theta_{j_3}^{t+1}$};
		\draw[->] (1) -- (2);
		\draw[->] (2) -- (3);
		\draw[->] (3) -- (1p);
		\draw[->] (1p) -- (2p);
		\draw[->] (2p) -- (3p);
	\end{tikzpicture}
	\caption {\textit{Random ordering of the conditional updating steps when comparing three models: $(i_1,i_2,i_3)$ and $(j_1,j_2,j_3)$ are random permutations of $(1,2,3)$, $j_1\neq i_3$}.}
	\label{new_definition}
\end{figure}
This produces three Markov chains, $(\theta_1^t)$, $(\theta_2^t)$ and $(\theta_3^t)$, and leads to the following definition.
\begin{Def}
	The multiple-model-comparison integral priors $\pi_1(\theta_1)$, $\pi_2(\theta_2)$ and $\pi_3(\theta_3)$, are the $\sigma$-finite invariant measures for the Markov chains $(\theta_1^t)$, $(\theta_2^t)$ and $(\theta_3^t)$, respectively,  when using a random ordering of the conditional updates.
\end{Def}
This definition reproduces the original one (with two models) and naturally extends to any number of models.

\section{Proper integral priors}

\subsection{Auxiliary compactified models}
As indicated above, for some model comparisons involving two models, the Markov chains associated with integral priors are recurrent Markov chains. In the general case \cite{Cano2013} have proposed to consider constrained imaginary training samples $z$ when the simulation of the chains is performed, which guarantees both the existence and the uniqueness of invariant probability measures. In brief, their practical recommendation consists in keeping the imaginary training samples constrained within an interval of $\pm 5s$ about the sample mean, where $s$ is the sample standard deviation. This approach has been implemented successfully in several situations comparing two models, see \cite{Cano2013} and \cite{Cano2018}. However, since the simulation of constrained imaginary training samples is carried out using some accept-reject method (for instance, a truncated normal distribution (\citealp{Robert1995})), this induces a consequent computational overhead. The computational alternative developed in this article consists in including in the comparison an artificial copy of each model of interest but with a corresponding artificial compact parametric space.

Consider for instance the case of two models under comparison, $M_i: f_i(x\mid\theta_i)$, $\theta_i\in\Theta_i$, $i=1,2$, and their attached improper priors $\pi_1^N(\theta_1)$ and $\pi_2^N(\theta_2)$, respectively. We then introduce an auxiliary model $M_3$ that is a copy of model $M_1$ with a parameter space restricted to a compact, that is, 
\[
M_3: f_{1}(x\mid\theta_3),\,\,\,\theta_3\in\Theta_3,
\]
where $\Theta_3$ is an artificial compact subset of $\Theta_1$, hence  $\pi_3^{N}(\theta_3)\propto\pi_1^N(\theta_3)\mathbb I_{\Theta_3}(\theta_3)$ is a proper prior. Our goal is to obtain integral priors for both $\theta_1$ and $\theta_2$, but we can resort to our generalized methodology of constructing integral priors for the three models $\{M_1,M_2,M_3\}$, which produces Markov chains $(\theta_i^t)$ with transition densities $Q_i(\theta_i\mid\theta_i^{\prime})$, and resulting integral priors $\pi_i(\theta_i)$, that is,
\[
\pi_i(\theta_i)=\int Q_i(\theta_i\mid\theta_i^{\prime})\pi_i(\theta_i^{\prime})\mathrm{d}\theta_i^{\prime},\,\,\,i=1,2,3.
\] 
Note that we are using the conditional notation for the marginal transition density $Q_i(\theta_i\mid\theta_i^{\prime})$, while the alternative notation $\pi_{ij}(\theta_i,\theta_j)$ corresponds for the between-model transition $\theta_i\rightarrow\theta_j$, $i\neq j$.

Since $\pi_3^{N}(\theta_3)$ is a proper prior, we need no imaginary training sample for $\theta_3$. In this case
\[
\pi_{13}(\theta_1,\theta_3)=\pi_{23}(\theta_2,\theta_3)=\pi_3^{N}(\theta_3).
\]
Thus, the Markov chain $(\theta_3^t)$ is a sequence of $\mathrm{i.i.d.}$ random variables, $\theta_3^t\sim\pi_3^{N}(\theta_3)$, and therefore $\pi_3^{N}(\theta_3)$ is the integral prior for $\theta_3$.
\begin{figure}[!t]
	\centering
	\begin{tikzpicture}[->,>= stealth',shorten >=2pt,line width =0.5 pt,node distance =2 cm,style ={minimum size=5mm},bend left=12]
		\node [] (1) at (0, 4) {$\theta_{i_1}^t$};
		\node [] (2) at (0, 2) {$\theta_{i_2}^t$};
		\node [] (3) at (0, 0) {$\theta_{i_3}^t$};
		\node [] (1p) at (3, 4) {$\theta_{j_1}^{t+1}$};
		\node [] (2p) at (3, 2) {$\theta_{j_2}^{t+1}$};
		\node [] (3p) at (3, 0) {$\theta_{j_3}^{t+1}$};
		\node [] (1pp) at (6, 4) {$\theta_{k_1}^{t+2}$};
		\node [] (2pp) at (6, 2) {$\theta_{k_2}^{t+2}$};
		\node [] (3pp) at (6, 0) {$\theta_{k_3}^{t+2}$};
		\draw[->] (1) -- (2);
		\draw[->] (2) -- (3);
		\draw[->] (3) -- (1p);
		\draw[->] (1p) -- (2p);
		\draw[->] (2p) -- (3p);
		\draw[->] (3p) -- (1pp);
		\draw[->] (1pp) -- (2pp);
		\draw[->] (2pp) -- (3pp);
	\end{tikzpicture}
	\caption {\textit{Illustration of a 2-step transition when comparing three models: $(i_1,i_2,i_3)$, $(j_1,j_2,j_3)$, and $(k_1,k_2,k_3)$ are random permutations of $(1,2,3)$, $i_3\neq j_1$, $j_3\neq k_1$}.}
	\label{2step_new_definition}
\end{figure}
The study of the stochastic stability of the corresponding Markov chains $(\theta_1^t)$ and $(\theta_2^t)$ entails a slightly greater level of complexity. The most natural approach is based on the 2-step transition densities
\[
Q_i^2(\theta_i\mid\theta_i^{\prime})=\int Q_i(\theta_i\mid\tilde{\theta_i}) Q_i(\tilde{\theta}_i\mid\theta_i^{\prime})\mathrm{d}\tilde{\theta}_i,\,\,\,i=1,2.
\]
For $i=1,2$, both the transition density $Q_i(\theta_i\mid\theta_i^{\prime})$ and the 2-step transition density $Q_i^2(\theta_i\mid\theta_i^{\prime})$, are mixtures of different conditional densities due to the randomness in the order of the between-model sequences, as illustrated by Figure \ref{2step_new_definition}. Each component of the mixture $Q_i^2(\theta_i\mid\theta_i^{\prime})$ corresponds to a specific realisation of the sequences $(i_1,i_2,i_3)$, $(j_1,j_2,j_3)$, and $(k_1,k_2,k_3)$. However, all components of this mixture share the crucial property that a simulation from $\pi_3^N(\theta_3)$ occurs within the sequence,  since $j=3$ is necessarily part of the sequence $(j_1,j_2,j_3)$. For instance, with regard to the mixture $Q_1^2(\theta_1\mid\theta_1^{\prime})$, configuration 1 in Figure \ref{2step_new_definition_example} corresponds to the component
\[
\int\pi_3^N(\theta_3)\pi_{31}(\theta_3,\theta_1)\mathrm{d}\theta_3,
\]
and configuration 2 corresponds to the component
\[
\int \pi_3^N(\theta_3)\pi_{32}(\theta_3,\theta_2)\pi_{21}(\theta_2,\tilde{\theta}_1)\pi_{12}(\tilde{\theta}_1,\theta_2^{\prime})\pi_{21}(\theta_2^{\prime},\theta_1)\mathrm{d}\theta_2^{\prime}\mathrm{d}\tilde{\theta}_1\mathrm{d}\theta_2\mathrm{d}\theta_3.
\]

\begin{figure}[!htb]
	\centering
	\begin{tikzpicture}[->,>= stealth',shorten >=2pt,line width =0.5 pt,node distance =2 cm,style ={minimum size=5mm},bend left=12]
		\node [] (0) at (2,5) {Configuration 1};
		\node [] (0q) at (8,5) {Configuration 2};
		\node [] (1) at (0, 4) {$\theta_{3}^t$};
		\node [] (2) at (0, 2) {$\theta_{2}^t$};
		\node [] (3) at (0, 0) {$\theta_{1}^t$};
		\node [] (1p) at (2, 4) {$\theta_{2}^{t+1}$};
		\node [] (2p) at (2, 2) {$\theta_{3}^{t+1}$};
		\node [] (3p) at (2, 0) {$\theta_{1}^{t+1}$};
		\node [] (1pp) at (4, 4) {$\theta_{3}^{t+2}$};
		\node [] (2pp) at (4, 2) {$\theta_{1}^{t+2}$};
		\node [] (3pp) at (4, 0) {$\theta_{2}^{t+2}$};
		\node [] (1q) at (6, 4) {$\theta_{3}^t$};
		\node [] (2q) at (6, 2) {$\theta_{2}^t$};
		\node [] (3q) at (6, 0) {$\theta_{1}^t$};
		\node [] (1pq) at (8, 4) {$\theta_{3}^{t+1}$};
		\node [] (2pq) at (8, 2) {$\theta_{2}^{t+1}$};
		\node [] (3pq) at (8, 0) {$\theta_{1}^{t+1}$};
		\node [] (1ppq) at (10, 4) {$\theta_{2}^{t+2}$};
		\node [] (2ppq) at (10, 2) {$\theta_{1}^{t+2}$};
		\node [] (3ppq) at (10, 0) {$\theta_{3}^{t+2}$};
		\draw[->] (1) -- (2);
		\draw[->] (2) -- (3);
		\draw[->] (3) -- (1p);
		\draw[->] (1p) -- (2p);
		\draw[->] (2p) -- (3p);
		\draw[->] (3p) -- (1pp);
		\draw[->] (1pp) -- (2pp);
		\draw[->] (2pp) -- (3pp);
		\draw[->] (1q) -- (2q);
		\draw[->] (2q) -- (3q);
		\draw[->] (3q) -- (1pq);
		\draw[->] (1pq) -- (2pq);
		\draw[->] (2pq) -- (3pq);
		\draw[->] (3pq) -- (1ppq);
		\draw[->] (1ppq) -- (2ppq);
		\draw[->] (2ppq) -- (3ppq);
	\end{tikzpicture}
	\caption {\textit{Two realizations of the  2-step transition process given in Figure \ref{2step_new_definition}}.}
	\label{2step_new_definition_example}
\end{figure}

More precisely, it is straightforward to derive that all components of these mixtures are necessarily equal to one of four cases that can be schematically represented as follows
\[
1\rightarrow 3\rightarrow 1
,\,\,\,
1\rightarrow 3\rightarrow 2\rightarrow 1,
\]
\[
1\rightarrow 3\rightarrow 1\rightarrow 2\rightarrow 1,\,\,\,\mathrm{and}\,\,\,
1\rightarrow 3\rightarrow 2\rightarrow 1\rightarrow 2\rightarrow 1.
\]
Then $Q^2_1(\theta_1\mid\theta_1^{\prime})$ does not depend on $\theta_1^{\prime}$, and hence $\pi_1(\theta_1)=Q^2_1(\theta_1\mid\theta_1^{\prime})$ is the integral prior for $\theta_1$:
\[
\int\pi_i(\theta_i)Q_i(\theta_i^*\mid\theta_i)\mathrm{d}\theta_i=\int Q_i^2(\theta_i\mid\theta_i^{\prime})Q_i(\theta_i^*\mid\theta_i)\mathrm{d}\theta_i
\]
\[
=\int Q_i(\theta_i\mid\tilde{\theta}_i)Q_i(\tilde{\theta}_i\mid\theta_i^{\prime})Q_i(\theta_i^*\mid\theta_i)\mathrm{d}\theta_i\mathrm{d}\tilde{\theta}_i
\]
\[
=\int Q_i(\tilde{\theta}_i\mid\theta_i^{\prime})Q_i^2(\theta_i^*\mid \tilde{\theta}_i)\mathrm{d}\tilde{\theta}_i=Q_i^2(\theta_i^*\mid \theta_i^{\prime})=\pi_i(\theta_i^*).
\]
Similarly, the 2-step transition for $(\theta_2^t)$ is the integral prior for $\theta_2$, that is,  $\pi_2(\theta_2)=Q^2_2(\theta_2\mid\theta_2^{\prime})$. From a practical perspective, we point out that all simulation steps that precede the latest simulation of $\theta_3$ in $Q_i(\theta_i\mid\theta_i^\prime)$ are superfluous since $\theta_3$ is generated independently.

However, these resulting integral priors do depend on which model is copied into a compactified version. Towards the elimination of the resulting arbitrariness, we propose to duplicate {\em each} model into a compactified version. For instance, when facing two models, we add to model $M_2$ the model
\[
M_4: f_{2}(x\mid\theta_4),\,\,\,\theta_4\in\Theta_4,
\]
where $\Theta_4$ is an artificial compact subset of $\Theta_2$ with $\pi_4^{N}(\theta_4)\propto\pi_2^N(\theta_4)\mathbb I_{\Theta_4}(\theta_4)$ thus a proper prior. While one could consider comparing $\{M_3,M_4\}$ instead of $\{M_1,M_2\}$, since the corresponding Bayes factor $B_{34}(x)$ is then well-defined, the models of interest remain $M_1$ and $M_2$. In fact, in the case where $M_1$ corresponds to the point null hypothesis $\theta_1=\theta_1^*$, the Bayes factor for the comparison $M_3$ \textit{versus} $M_4$ is
\[
B_{34}(x)=\frac{f_1(x\mid\theta_1^*)\int_{\Theta_4} \pi_2^N(\theta_2)\mathrm{d}\theta_2}{\int_{\Theta_4} f_2(x\mid\theta_2)\pi_2^N(\theta_2)\mathrm{d}\theta_2},
\]
which typically goes to +$\infty$ as $\Theta_4$ approaches $\Theta_2$. Therefore moving to the set of models $\{M_1,M_2,M_3,M_4\}$ is purely procedural and solely intended to derive a well-defined integral prior on both $M_1$ and $M_2$.

For the same reason as above, the training sample size for $\theta_4$ is zero and the integral prior is again $\pi_4^N(\theta_4)$.
The 2-step transition $Q^2_i(\theta_i\mid\theta_i^{\prime})$ does not depend on $\theta_i^{\prime}$ and hence $\pi_i(\theta_i)=Q^2_i(\theta_i\mid\theta_i^{\prime})$ is the integral prior for $\theta_i$, $i=1,2$. Besides, $Q^2_1(\theta_1\mid\theta_1^{\prime})$ is a mixture of densities and the components of this mixture are reduced to eight cases that can be schematically represented as follows:
\[
1\rightarrow 3\rightarrow 1
,\,\,\,
1\rightarrow 3\rightarrow 2\rightarrow 1,
\]
\[
1\rightarrow 3\rightarrow 1\rightarrow 2\rightarrow 1,\,\,\,
1\rightarrow 3\rightarrow 2\rightarrow 1\rightarrow 2\rightarrow 1,
\]
\[
1\rightarrow 4\rightarrow 1
,\,\,\,
1\rightarrow 4\rightarrow 2\rightarrow 1,
\]
\[
1\rightarrow 4\rightarrow 1\rightarrow 2\rightarrow 1,\,\,\,\mathrm{and}\,\,\,
1\rightarrow 4\rightarrow 2\rightarrow 1\rightarrow 2\rightarrow 1,
\]
which means that there exist functions $H_1^1(\theta_1,\theta_3)$ and $H_1^2(\theta_1,\theta_4)$, defined for $\theta_1,\theta_3\in\Theta_1$ and $\theta_4\in\Theta_2$, such that the integral prior $\pi_1(\theta_1)$ is
\[
\pi_1(\theta_1)=Q_1^2(\theta_1\mid\theta_1^{\prime})=\int\pi_3^N(\theta_3)H_1^1(\theta_1,\theta_3)\mathrm{d}\theta_3+\int\pi_4^N(\theta_4)H_1^2(\theta_1,\theta_4)\mathrm{d}\theta_4.
\]
Similarly, there exist functions $H_2^1(\theta_2,\theta_3)$ and $H_2^2(\theta_2,\theta_4)$, defined for $\theta_2,\theta_4\in\Theta_2$ and $\theta_3\in\Theta_1$, such that
\[
\pi_2(\theta_2)=Q_2^2(\theta_2\mid\theta_2^{\prime})=\int\pi_3^N(\theta_3)H_2^1(\theta_2,\theta_3)\mathrm{d}\theta_3+\int\pi_4^N(\theta_4)H_2^2(\theta_2,\theta_4)\mathrm{d}\theta_4,
\]
is the integral prior for $\theta_2$.

Obviously, this device of adding copycat models with an artificial compact parametric space applies to more than two models. We start from $q\geq 2$ models
\[
M_i: f_i(x\mid\theta_i),\,\,\,\theta_i\in\Theta_i,\,\,\,i=1,\dots,q,
\]
under comparison and associated improper prior distributions $\pi_i^N(\theta_i)$ $i=1,\dots,q$. Given a compact subset $\Theta_{q+i}\subseteq\Theta_i$ such that $\int_{\Theta_{q+i}} \pi_i^N(\theta_i)\mathrm{d}\theta_i<+\infty$, we consider a copy of  $M_i$ but with parameter space $\Theta_{q+i}$, that is,
\[
M_{q+i}:f_i(x\mid\theta_{q+i}),\,\,\,\theta_{q+i}\in\Theta_{q+i},
\]
and $\pi_{q+i}^N(\theta_{q+i})\propto\pi_i^N(\theta_{q+i})\mathbb I_{\Theta_{q+i}}(\theta_{q+i})$, $i=1,\dots,q$. Then integral priors for $\{M_1,\dots,M_q,$ $\dots,M_{2q}\}$ do exist, they all are unique and proper priors, and the integral prior for $\theta_i$ satisfies 
\[
\pi_i(\theta_i)=Q_i^2(\theta_i\mid\theta_i^{\prime})=\sum_{j=1}^q\int\pi_{q+j}^N(\theta_{q+j})H_i^j(\theta_i,\theta_{q+j})\mathrm{d}\theta_{q+j},
\]
where the function $H_i^j(\theta_i,\theta_{q+j})$ is defined for $\theta_i\in\Theta_i$ and $\theta_{q+j}\in\Theta_j$, $i,j=$ $1,\dots,q$. 

\subsection{Selecting the compact sets}
Without loss of generality consider two models under comparison, $M_1$ and $M_2$, with their copies over compact parametric spaces, $M_3$ and $M_4$, respectively. It is quite sensible to seek compact sets such that the posterior probabilities
\[
P_x^3=\int_{\Theta_3}\pi_1^N(\theta_1\mid x)\mathrm{d}\theta_1\,\,\,\mathrm{and}\,\,\,P_x^4=\int_{\Theta_4}\pi_2^N(\theta_2\mid x)\mathrm{d}\theta_2\,
\]
are close to one. While there are a variety of ways to choose these compact sets, we opt for the simple solution where both $\Theta_3$ and $\Theta_4$ are the Cartesian products of compact intervals based on marginal posterior quantiles in such a way that the compact sets include the highest posterior values. Note that here the posterior distributions are associated with the original prior distributions, independently for each model. More specifically, our proposal is made of credible intervals with posterior probability $1-\alpha$, based on the quantiles $\alpha/2$ and $1-\alpha/2$, with $\alpha\in(0,1)$ small, typically $\alpha\leq 0.05$. Note that the computation of such HPD regions usually requires MCMC samples. This differs from the costlier requirement of simulating constrained imaginary training samples in the Markov chains that define integral priors under the approach proposed in \cite{Cano2013}. The following example illustrates the behaviour of this method.

\begin{example}
	For the data $x=(x_1,\dots,x_n)$, consider the models
	\[
	M_1:f_1(x_j\mid \theta_1)=\frac{1}{\theta_1}\exp(-x_j/\theta_1),\,\,\,j=1,\dots,n,\,\,\,\theta_1\in(0,1),
	\]
	and
	\[
	M_2:f_2(x_j\mid \theta_2)=\frac{1}{\theta_2}\exp(-x_j/\theta_2),\,\,\,j=1,\dots,n,\,\,\,\theta_2\in(1,+\infty),
	\]
	with $\pi_i^N(\theta_i)= 1/\theta_i$, $i=1,2$. Note that $\sum_j x_j$ is a sufficient statistic for $\theta$, and that given prior distributions for $\theta_i$, $i=1,2$, the Bayes factor is determined by $n$ and $\bar{x}$. For a sample (see supplementary material) of size $n=15$ and mean $0.7$, we implement the above methodology of integral priors, thus defining the collection of models $\{M_1,M_2,M_3,M_4\}$, where $\Theta_{i+2}=[a_{i+2},b_{i+2}]$ is a $100(1-\alpha)$\% credible interval based on the quantiles $\alpha/2$ and $1-\alpha/2$ of $\pi_i^N(\theta_i\mid x)$, $i=1,2$. We also opt for using training samples $z$ of size one for both $\theta_1$ and $\theta_2$. Simulation from the posterior distributions is straightforward:
	\[
	\theta_1=1/(1-z^{-1}\log u_1),\,\,\,u_1\sim\mathcal{U}(0,1),
	\]
	and
	\[
	\theta_2=-z/\log(u_2(1-e^{-z})+e^{-z}),\,\,\,u_2\sim\mathcal{U}(0,1).
	\]
	For $\alpha\in\{0.05,0.01,0.005,0.001\}$, Table \ref{table1} provides $\hat{B}_{12}(x)$$=1/\hat{B}_{21}(x)$ where $\hat{B}_{21}(x)$ is given by (\ref{MCestimator}), the approximate values of $B_{12}(x)$, as well as the estimated posterior probability of model $M_1$ if the prior probabilities of $M_1$, $M_2$, $M_3$, and $M_4$ are $1/2$, $1/2$, $0$, and $0$, respectively. Hence, $\hat{P}(M_1\mid x)=\hat{B}_{12}(x)/(1+\hat{B}_{12}(x))$. 
	
		\begin{table}[t]
		\begin{center}
			\caption{\textit{Estimation of the Bayes factor $B_{12}(x)$ and posterior probability $P(M_1\mid x)$, based on $10^6$ simulations from the integral priors}.}		\label{table1}
		\begin{tabular}{lcccccc}
			\hline
			\rule{0pt}{14pt}	$\alpha$ & $a_3$    & $b_3$    & $a_4$    & $b_4$    & $\hat{B}_{12}(x)$   & $\hat{P}(M_1\mid x)$ \\
			\hline
			\hline
			0.05  & 0.44  & 0.97  & 1.00  & 1.62  & 8.98  & 0.90  \\
			0.01  & 0.39  & 0.99  & 1.00  & 1.90  & 9.10  & 0.90  \\
			0.005 & 0.37  & 1.00  & 1.00  & 2.02  & 9.09  & 0.90  \\
			0.001 & 0.34  & 1.00  & 1.00  & 2.28  & 9.16  & 0.90  \\
			\hline
		\end{tabular}
	\end{center}

	\end{table}

It is worth noting that under the unconstrained Exponential model
\[
f(x_j\mid\theta)=\theta^{-1}\exp(-x_j/\theta),\,\,\,j=1\dots,n,\,\,\,\theta\in(0,+\infty),
\]
and $\pi^N(\theta)\propto 1/\theta$, the posterior probability of $\Theta_1=(0,1)$ is 0.89.

The histograms corresponding to the chains $(\log \theta_i^t)$, $i=1,2,3,4$, using the associated Markov chains with the compact sets for $\alpha=0.001$ (last row in Table \ref{table1}), are shown in Figure \ref{fig:exponentialexample1}.
\begin{figure}
	\centering
	\includegraphics[width=0.7\linewidth]{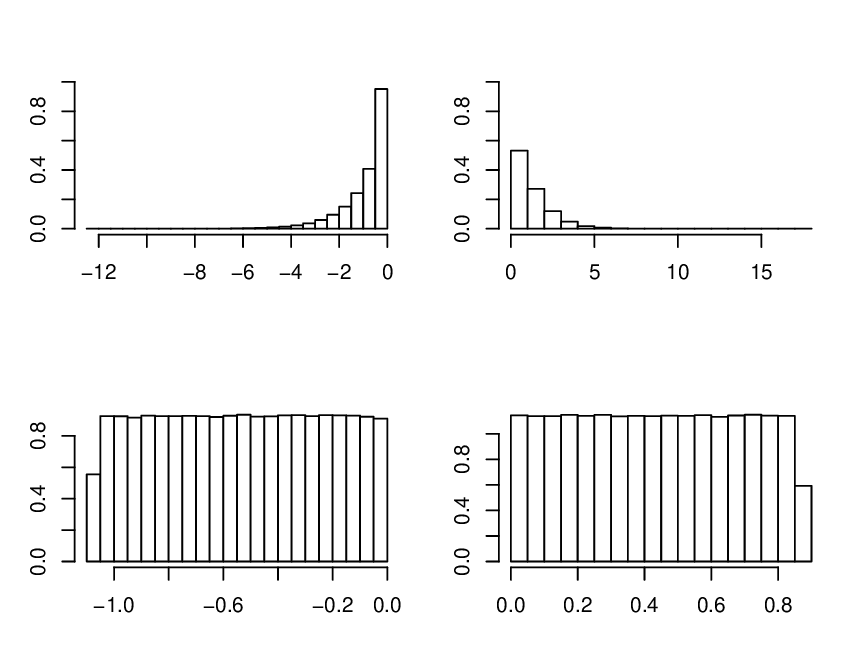}
	\caption{\textit{Histograms of $10^6$ simulations from the integral priors for $\log \theta_i$, $i = 1, 2$ (first row: $\log\theta_1$ left,  $\log\theta_2$ right), and $i=3, 4$ (second row: $\log\theta_3$ left,  $\log\theta_4$ right)}.}
	\label{fig:exponentialexample1}
\end{figure}
	
To evaluate the Monte Carlo variability in Table \ref{table1}, we repeated 100 times the estimations, that is, we simulated the Markov chains 100 times and computed the corresponding estimator $\hat{B}_{12}(x)$. The standard deviation of the 100 values of $\hat{B}_{12}(x)$ was 0.02 for each value of $\alpha$.

%\cite{Moreno2005} has stated that the \textit{encompassing} approach for intrinsic priors does not work for this example, and has proposed an \textit{ad hoc} solution inspired in intrinsic priors for two different models: the point null hypothesis $\theta=1$ and the unconstrained model $\theta>0$. Then the proposal is to consider the restriction of the prior for the unconstrained model to $\theta<1$ and $\theta>1$ to obtain two priors for the computation of the Bayes factor.

The intrinsic priors for the arithmetic intrinsic Bayes factor do not exist for this example because the expectation of $B_{12}^N(z)$ is not finite. The encompassing approach for intrinsic priors has been suggested in such situations. This approach proposes embedding $M_1$ and $M_2$ in the larger model $M_0:$ $\theta\in(0,+\infty)$, and then to obtain intrinsic priors for nested models to compute $B_{20}^I(x)$ and $B_{01}^I(x)$, and finally $B_{21}^I(x)$ is defined as $B_{20}^I(x)B_{01}^I(x)$. \cite{Moreno2005} has stated that this approach does not work for this example.
	
Instead, \cite{Moreno2005} has proposed an ad hoc solution inspired from intrinsic priors for two different models: the point null hypothesis $\theta=1$ and the unconstrained model $\theta > 0$. Then the proposal is to consider the restriction of the intrinsic prior for the unconstrained model to $\theta < 1$ and $\theta > 1$ to obtain two priors for the computation of the Bayes factor  $\tilde{B}_{12}^I(x)$. For a sample $x$ of size $n = 15$ and mean 0.7, this Bayes factor is $\tilde{B}_{12}^I(x)=7.6$, which is similar to the values in Table \ref{table1}. \cite{BergerMortera1999} have obtained other priors for the comparison $M_1$ versus $M_2$; for example, the intrinsic priors for the median intrinsic Bayes factor and for the fractional Bayes factor. With these priors the resulting Bayes factors are $\tilde{B}_{12}^{MIBF}(x)=5.6$ and $\tilde{B}_{12}^{F}(x)=3.1$, respectively. Both values are significantly different from those obtained with our approach. This is partly explained by the fact that $\int_1^{+\infty} \pi^I_2(\theta_2)\mathrm{d}\theta_2/\int_0^1 \pi^I_1(\theta_1)\mathrm{d}\theta_1$ is 2.67 for the priors used in $\tilde{B}_{12}^{F}(x)$, and 1.46 for the priors used in $\tilde{B}_{12}^{MIBF}(x)$,  and this translates into an a priori advantage for model $M_2$, as pointed out in \cite{BergerMortera1999}.
\end{example}

\section{Test on a Normal mean under dependent samples}\label{testNormalmean}
In the specific context of paired samples in which each subject is measured twice, resulting in pairs of observations, there are three natural hypotheses regarding the mean difference, $\mu$, namely, $\mu<0$, $\mu=0$, and $\mu>0$. This multiple non-nested hypothesis testing problem has been previously studied in \cite{BergerMortera1999}, assuming known variance and comparing  the encompassing arithmetic intrinsic Bayes factor (EIBF), the encompassing expected intrinsic Bayes factor (EEIBF), the median intrinsic Bayes factor, and the fractional Bayes factor. Unfortunately, these approaches do not correspond to Bayes factors with respect to genuine prior distributions (see \cite{BergerMortera1999}, p552). In this section, we therefore apply the theory of integral priors for this testing problem. 

Under model $M_i$ the difference data $x=(x_1,\dots,x_n)$ are independently drawn from the Normal distribution $\mathcal{N}(\mu_i,\sigma_i^2)$, $i=1,2,3$, with $\mu_1<0$, $\mu_2=0$, and $\mu_3>0$, respectively, that is,
\[
f_i(x\mid\mu_i,\sigma_i)=(2\pi)^{-n/2}\sigma_i^{-n}\exp\left(-\frac{1}{2\sigma_i^2}\sum_{j=1}^n(x_j-\mu_i)^2\right),\,\,\,i=1,3,
\]
\[
f_2(x\mid\sigma_2)=(2\pi)^{-n/2}\sigma_2^{-n}\exp\left(-\frac{1}{2\sigma_2^2}\sum_{j=1}^nx_j^2\right),
\]
and we consider the default estimation priors $\pi_i^N(\theta_i)= 1/\sigma_i$, $i=1,2,3$, where $\theta_1=(\mu_1,\sigma_1)$, $\theta_2=\sigma_2$, and $\theta_3=(\mu_3,\sigma_3)$.

To investigate this setting, we produced a simulated sample $x$ of size $n=15$ from the Normal distribution with mean $-1$ and variance $4$. The observed sample mean and standard deviation were $-0.927$ and $2.530$, respectively. We thus consider copies $M_4$, $M_5$, and $M_6$, with compact parametric spaces $\Theta_i$, $i=4,5,6$, of the form
\[
\theta_4=(\mu_4,\sigma_4)\in \Theta_4= [a_4,b_4]\times [c_4,d_4],
\]
\[
\theta_5=\sigma_5 \in \Theta_5= [c_5,d_5],
\]
and
\[
\theta_6=(\mu_6,\sigma_6)\in \Theta_6= [a_6,b_6]\times [c_6,d_6],
\]
respectively. 
%and the methodology of integral priors for $\{M_1,\dots, M_6\}$.

For models $M_1$ and $M_3$, imaginary training samples are samples, $z$, of size 2, and the corresponding posterior distributions are
\[
\pi_i^N(\mu_i,\sigma_i\mid z)\propto \sigma_i^{-3}\exp\left(-\frac{s_z^2}{2\sigma_i^2}\right)\exp\left(-\frac{(\bar{z}-\mu_i)^2}{\sigma_i^2}\right),\,\,\,i=1,3,
\]
where $\bar{z}$ and $s_z\neq 0$ are the sample mean and standard deviation of $z$, respectively. The posterior distribution $\pi_i^N(\mu_i\mid z)$ is proportional to
\[
\int_0^{+\infty}\sigma_i^{-3}\exp\left(-\frac{s_z^2}{2\sigma_i^2}\right)\exp\left(-\frac{(\bar{z}-\mu_i)^2}{\sigma_i^2}\right)\mathrm{d}
\sigma_i=\frac{1}{s_z^2+2(\bar{z}-\mu_i)^2}\,,
\]
an half-Cauchy density, and therefore the simulation from $\pi_i^N(\mu_i\mid z)$ can be carried out using the probability integral transform method, that is, we can generate $u,v\sim\mathcal{U}(0,1)$
and take
\[
\mu_1=\bar{z}-\frac{s_z}{\sqrt{2}}\cot\left(\frac{u}{2}(\pi-2\arctan(\sqrt{2}\bar{z}/s_z))\right)
\]
and
\[
\mu_3=\bar{z}-\frac{s_z}{\sqrt{2}}\cot\left(\frac{v-1}{2}(\pi+2\arctan(\sqrt{2}\bar{z}/s_z))\right).
\]

To simulate from $\pi_i^N(\sigma_i\mid \mu_i, z)$ we can generate $\xi_i$ from the Exponential distribution $\mathcal{E}xp(1)$ and take 
\[
\sigma_i=\sqrt{\frac{s_z^2+2(\bar{z}-\mu_i)^2}{2\xi_i}},\,\,\,i=1,3.
\]
For model $M_2$, imaginary training samples are samples of size 1, $z\neq 0$, and the posterior distribution is
\[
\pi_2^N(\sigma_2 \mid z)\propto \sigma_2^{-2}\exp\left(-\frac{z^2}{2\sigma_2^2}\right).
\]
Therefore, to simulate from $\pi_2^N(\sigma_2 \mid z)$ we can generate $\xi_2\sim\mathcal{G}a(1/2,1)$ and take $\sigma_2=|z|/\sqrt{2\xi_2}$.

For $\alpha\in\{0.05,0.01,0.005\}$ we again select the compact intervals as the $100(1-\alpha)$\% credible intervals based on the quantiles $\alpha/2$ and $1-\alpha/2$ of the marginal posterior distribution of $\pi_i^N(\theta_i\mid x)$, $i=1,2,3$, that is
\[
\alpha/2=\int_{-\infty}^{a_{i+3}}\pi_i^N(\mu_i\mid x)\mathrm{d}\mu_i,\,\,\,1-\alpha/2=\int_{-\infty}^{b_{i+3}}\pi_i^N(\mu_i\mid x)\mathrm{d}\mu_i\,\,\,i=1,3,
\]
\[
\alpha/2=\int_{0}^{c_{i+3}}\pi_i^N(\sigma_i\mid x)\mathrm{d}\sigma_i,\,\,\,\mathrm{and}\,\,\,1-\alpha/2=\int_{0}^{d_{i+3}}\pi_i^N(\sigma_i\mid x)\mathrm{d}\sigma_i\,\,\,i=1,2,3.
\]

Based on such compact sets, we apply the methodology of integral priors with $500,000$ simulations in order to approximate the posterior probabilities of models $M_1$, $M_2$ and $M_3$ when selecting as set of prior probabilities $\{1/3, 1/3, 1/3, 0, 0, 0\}$. The approximated posterior probabilities of models $M_1$, $M_2$, and $M_3$ are shown in Table \ref{table2}, again with very little variability when $\alpha$ decreases. The execution time was 20 seconds for each row in Table \ref{table2}. On the other hand, we have applied integral priors for $\{M_1,M_2,M_3\}$ with constrained imaginary training samples as proposed in \cite{Cano2013} with Markov chains of length 500,000. As could be expected, the execution time was higher with this approach. Specifically, the execution time was 120 seconds and the posterior probabilities of models $M_1$, $M_2$, and $M_3$ were 0.52, 0.46, and 0.02, respectively.

For this example we have computed the fractional Bayes factors and the median intrinsic Bayes factors. The corresponding posterior probabilities of models $\{M_1, M_2,M_3\}$ are $\{0.41, 0.53, 0.06\}$ and $\{0.28, 0.63, 0.09\}$, respectively. Note that the conclusion from these methodologies would be to choose $M_2$, while the conclusion based on integral priors would be to choose $M_1$. However, despite this slight numerical difference, the posterior probability is not sufficiently conclusive in favor of either $M_1$ or $M_2$ in any case. In fact, although the fractional Bayes factor (0.77) and the median intrinsic Bayes factor (0.44) provide evidence against $M_1$, and the Bayes factor for integral priors (0.88) provides evidence against $M_2$, all of these pieces of evidence are not worth more than a bare mention, according to Jeffreys’ scale of evidence.

\begin{figure}
	\centering
	\includegraphics[width=0.72\linewidth]{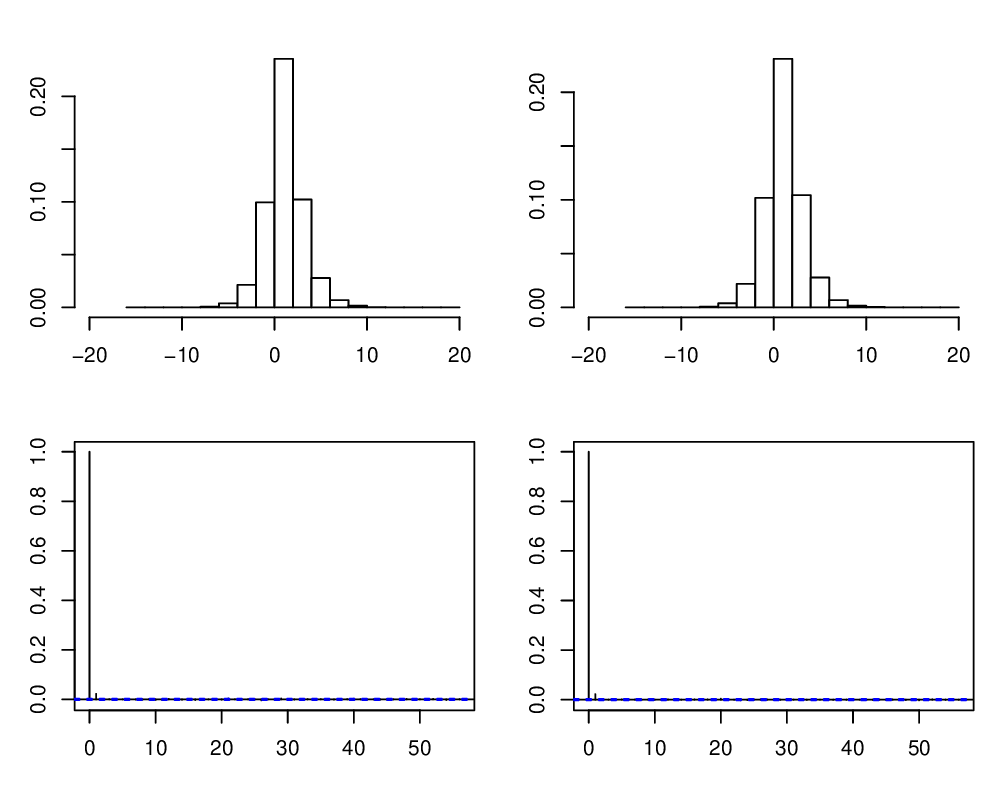}
	\caption{\textit{Histograms and empirical autocorrelations based on simulations from the integral priors for $\log(-\mu_1)$ and $\log(\mu_3)$ in the Normal mean difference example}.}
	\label{dependetsmeansposteriorsmh}
\end{figure}

Figure \ref{dependetsmeansposteriorsmh} shows the histograms and the  empirical autocorrelations based on the simulations from the integral prior for $\log(-\mu_1)$ and $\log(\mu_3)$, corresponding to $\alpha=0.005$.

\begin{table}[!t]
	\begin{center}
			\caption{\textit{Estimation of the posterior probabilities of models $M_1$, $M_2$, and $M_3$ in the Normal mean difference example}.}	\label{table2}
	\begin{tabular}{lccc}
		\hline
		\rule{0pt}{14pt}	$\alpha$ & $\hat{P}(M_1\mid x)$ & $\hat{P}(M_2\mid x)$ & $\hat{P}(M_3\mid x)$ \\
		\hline
		\hline
		0.05  & 0.51  & 0.42  & 0.06 \\
		0.01  & 0.50  & 0.43  & 0.06 \\
		0.005 & 0.50  & 0.44  & 0.06 \\
		\hline
	\end{tabular}
\end{center}

\end{table}

\section{Integral priors for ANOVA}

\subsection{Original and auxiliary models}
In the classical ANOVA setting, we now consider a sample $x_i$ from the Normal population $\mathcal{N}(\mu_i,\sigma^2)$, $i=1,\dots,k$, in order to handle on the hypothesis test
\[
\begin{array}{l}
	H_0:\,\mu_1=\dots =\mu_k\\
	H_1:\, \mu_i\, \mathrm{are\, not\, all\, equal}.
\end{array}
\]

Hence we are again dealing with a comparison between two models. Under model $M_1$ the $k$ populations have mean $\mu$ and variance $\sigma_1^2$, and under model $M_2$ the population $i$ has mean $\mu_i$ and variance $\sigma_2^2$, $i=1,\dots,k$. 
%Once we have selected prior distributions on the parameters, the Bayes factor is a key tool to perform model selection. 
The default (estimation) prior distributions in this setting are $\pi_1^N(\mu,\sigma_1)= 1/\sigma_1$ and $\pi_2^N(\mu_1,\dots,\mu_k,\sigma_2)= 1/\sigma_2$. Unfortunately, these priors are improper and therefore their use is not validated for model selection. In order to construct integral priors for ANOVA, we again consider copies $M_3$ and $M_4$ with compact parametric spaces.
%, and the methodology of integral priors for the models $\{M_1,M_2,M_3,M_4\}$.

%A key ingredient is the concept of imaginary training sample (ITS),  which is usually considered of minimal size, see \cite{BergerPericchi2004}. The ITS 
The imaginary training samples 
%for the transitions $\theta_1\rightarrow \theta_2$ and $\theta_2\rightarrow \theta_1$ 
are generated as follows:
\vspace{.5cm}

For $\theta_1\rightarrow \theta_2$. Given $\theta_1=(\mu,\sigma_1)$
\begin{enumerate}
	\item $j\sim\mathcal{U}\{1,\dots,k\}$
	\item $z_i\sim N(\mu,\sigma_1^2)$, $i\neq j$, and $z_j=(z_j^1,z_j^2)$, where $z_j^1$, $z_j^2\sim N(\mu,\sigma_1^2)$
	\item $z=(z_1,\dots,z_k)$ is the imaginary training sample
\end{enumerate}
\vspace{.5cm}

For $\theta_2\rightarrow \theta_1$. Given $\theta_2=(\mu_1,\dots,\mu_k,\sigma_2)$
\begin{enumerate}
	
	\item $j_1,j_2\sim\mathcal{U}\{1,\dots,k\}$
	\item $z_{j_1}\sim N(\mu_{j_1},\sigma_2^2)$, $z_{j_2}\sim N(\mu_{j_2},\sigma_2^2)$
	\item $z=(z_{j_1},z_{j_2})$ is the imaginary training sample
\end{enumerate}

For model $M_1$, the sample size is $2$, that is $x=(z_1,z_2)$. The simulation from $\pi_1^N(\mu,\sigma_1\mid x)$ can be carried out as follows:

\begin{enumerate}
	\item $\xi\sim\mathcal{G}a(1/2,1)$, $\sigma_1=\frac{\mid z_1-z_2\mid}{2\sqrt{\xi}}$
	\item $\mu\sim N(\overline{x},\sigma_1^2/2)$.
\end{enumerate}

For model $M_2$, the sample size is $n_j=2$ for some $j\in\{i,\dots,k\}$ and $n_i=1$ for $i\neq j$. Then
\[
\pi_2^N(\sigma_2\mid x)\propto \sigma_2^{-2}\exp\left(-\frac{s_j^2}{2\sigma_2^2} \right),
\]
and therefore the simulation from $\pi_2^N(\mu_1,\dots,\mu_k,\sigma_2\mid x)$ can be carried out as follows:

\begin{enumerate}
	\item $\xi\sim\mathcal{G}a(1/2,1)$, $\sigma_2=\frac{\mid z_1-z_2\mid}{2\sqrt{\xi}}$, where $x_j=(z_1,z_2)$
	\item $\mu_i\sim N(\overline{x}_i,\sigma_2^2/n_i)$, $i=1,\dots,k$.
\end{enumerate}

\subsection{Example}
To illustrate our methodology we consider $k=3$ populations. The compact parametric spaces are chosen as Cartesian products
\[
\Theta_3=[a_3,b_3]\times[c_3,d_3]
\]
and
\[
\Theta_4=[a_{41},b_{41}]\times[a_{42},b_{42}]\times[a_{43},b_{43}]\times[c_{4},d_{4}].
\]

For $\alpha\in\{0.05,0.01$,0.005$\}$, we select the intervals as $100(1-\alpha)\%$ credible intervals. For sample sizes $n_1=n_2=n_3=10$, we have considered sample means $\overline{x}_1=\overline{x}_2=0$, $\overline{x}_3\in\{0,0.75,1,1.5\}$, and sample standard deviations $s_1=s_2=s_3=1$. As in previous examples, the compact parametric spaces are based on the posterior quantiles $\alpha/2$ and $1-\alpha/2$ of the marginal posterior distribution obtained with $\pi_i^N (\theta_i)$, $i = 1, 2$.
Table \ref{tablaANOVA} produces estimated posterior probability $\hat{P}(M_1\mid x)$ under the prior probabilities $P(M_1)=P(M_2)=1/2$ and $P(M_3)=P(M_4)=0$.

%\begin{table}[htbp]
%	\begin{center}
%	\begin{tabular}{c|cccc}
%		\hline
%		& $\overline{x}_3$ &  $P_x^3$ & $P_x^4$ & $\hat{P}(M_1\mid x)$\\
%		\hline
%		\multirow{4}[2]{*}{$\alpha=0.05$} & 0        & 0.90  & 0.82  & 0.91   \\
%		& 0.75    & 0.90  & 0.82  & 0.67  \\
%		& 1         & 0.90  & 0.82  &  0.37 \\
%		& 1.5     & 0.90  & 0.82  &  0.03 \\
%		\hline
%		\multirow{4}[2]{*}{$\alpha=0.01$} & 0   & 0.98     & 0.96  & 0.91  \\
%		& 0.75    & 0.98  & 0.96 & 0.68   \\
%		& 1         &  0.98 &  0.96 & 0.39   \\
%		& 1.5     &  0.98 & 0.96  & 0.03   \\
%		\hline
%	\end{tabular}
% \end{center}
%	\caption{Estimations of the posterior probabilities of model $M_1$ in the ANOVA example.}
%	\label{tablaANOVA}
%\end{table}

\begin{table}[htbp]
	\begin{center}
		\caption{\textit{Estimations of the posterior probabilities of model $M_1$ in the ANOVA example}.}\label{tablaANOVA}
	\begin{tabular}{lcccc}
		\hline
		& \multicolumn{4}{c}{$\overline{x}_3$} \\
		\cline{2-5}
		& 0     & 0.75  & 1     & 1.5 \\
		\hline
		$\alpha=0.05$ & 0.93  & 0.70 & 0.38  & 0.03 \\
		$\alpha=0.01$ & 0.94  & 0.70 & 0.41  & 0.03 \\
		$\alpha=0.005$ & 0.94 & 0.71 & 0.41  & 0.03 \\
		\hline
	\end{tabular}
\end{center}

\end{table}

The $p$-values for the ANOVA test were 1, 0.173, 0.051, and 0.003, for $\overline{x}_3=0,0.75,1$, and $1.5$, respectively. Our reference priors are the $g$-priors as implemented in the R package {\sf BayesFactor} with the three possible options for the scale of the inverse-chi-square priors for the $g$'s: \textit{medium}, \textit{wide}, and \textit{ultrawide}. The posterior probability of model $M_1$ were 0.82 (\textit{medium}), 0.88 (\textit{wide}), and 0.93 (\textit{ultrawide}) when $\overline{x}_3=0$; 0.59 (\textit{medium}), 0.65 (\textit{wide}), and 0.73 (\textit{ultrawide}) when $\overline{x}_3=0.75$; 0.38 (\textit{medium}), 0.41 (\textit{wide}), and 0.50 (\textit{ultrawide}) when $\overline{x}_3=1$; and 0.06 (\textit{medium}), 0.06 (\textit{wide}), and 0.06 (\textit{ultrawide}) when $\overline{x}_3=1.5$.

\section{Poisson and negative binomial}

The article by \cite{Moreno2021} is dedicated to the test of a Poisson model {\em versus} the negative binomial family, using intrinsic priors. In this section we develop integral priors and compare the methodologies.

Let us consider the Poisson model
\[
M_1:\, f_1(x\mid \theta_1)=\frac{\theta_1^x e^{-\theta_1}}{x!},\,\,\,\theta_1>0,
\]
and the negative binomial models
\[ 
M_r:\, f_r(x\mid\theta_r)={x+r-2\choose x}\theta_r^{r-1}(1-\theta_r)^x, \,\,\,\theta_r\in(0,1),
\]
$r=2,\dots,q$. Jeffreys priors are $\pi_1^N(\theta_1)=\theta_1^{-1/2}$ and $\pi_r^N(\theta_r)=\theta_r^{-1}(1-\theta_r)^{-1/2}$, $r=2,\dots,q$, and the minimal training sample size is one in all cases. With these priors, the posteriors are Gamma and Beta distributions, respectively:
\[
\pi_1^N(\theta_1\mid z)=\mathcal{G}a\left(\theta_1\mid z+1/2,1\right)
\]
and
\[
\pi_r^N(\theta_r\mid z)=\mathcal{B}e(\theta_r\mid r-1,z+1/2),\,\,\,r\geq 2,
\]
for $z\geq 0$ a minimal training sample.

As is pointed out in \cite[][pp.1457-1458]{Moreno1998}, the intrinsic priors methodology is not well defined when comparing Poisson versus geometric ($M_2$) because of the arbitrariness in the limiting procedure for defining the intrinsic priors and the corresponding Bayes factor. This issue can be \textit{adjusted} if instead of using Jeffreys priors as the initial default estimation priors---which would be the natural choice---one uses Jaynes' priors, for which a minimal training sample is any value of $x$ other than $x=0$ 
%This has been the approach in  \cite{Moreno2021}.

Here we illustrate the integral priors methodology with the data in Table \ref{MadisonTABLA} referring the number of times that the word ``may" appears per block in papers by James Madison, see \cite{Coni2000}.
\begin{table}
	\begin{center}
		\caption{\textit{The word ``may" per block in papers by James Madison.}}\label{MadisonTABLA} 
		\begin{tabular}{|c|ccccccc|}
			\hline
			$X$ (number of occurrences)	& 0 & 1 & 2 & 3 & 4 & 5 & 6\\
			\hline
			Frequency (number of blocks) &  156 & 63 & 29 & 8 & 4 & 1 & 1\\
			\hline
		\end{tabular}
		
	\end{center}
\end{table}

We have considered $q=15$ and, as in previous examples, the compact parametric spaces are based on the quantiles $\alpha/2$ and $1-\alpha/2$ of the posterior distributions $\pi_i^N(\theta_i\mid y)$, $i=1,\dots,q$.

Based on such compact sets, we apply the methodology of integral priors with $50,000$ simulations in order to approximate the posterior probabilities of models $M_1,\dots,M_{15}$. The prior probabilities were $1/15$ for models $M_1,\dots,M_{15}$, and 0 for the artificial models. Using $\alpha=0.05$, the approximated posterior probabilities where 0.68 for the geometric model $M_2$, 0.26 for $M_3$, $0.04$ for $M_4$, and less than 0.01 for other models. The results were virtually identical with $\alpha = 0.001$. 

For the purpose of comparing the methodologies, we have applied the approach proposed in \cite{Moreno2021} obtaining quite similar values for the posterior probabilities: 0.70 for $M_2$, 0.24 for $M_3$, 0.04 for $M_4$, 0.01 and lower for the other models.

Furthermore, we have used simulated data to compare the models $\{M_1,M_2,M_3,$ $M_4,M_5\}$ using integral priors ($\alpha=0.05$, and chains of length 50,000) and the intrinsic priors developed in \cite{Moreno2021}. A sample of size $n$ is simulated from the \textit{True} model 100 times, and for each sample we have computed the posterior probabilities of the models, and the probability of the models based on Akaike weights. Table \ref{tab:addlabel} displays the means of these 100 probabilities. It can be observed that the values are very similar.

{\setlength{\tabcolsep}{3pt}
	\begin{table}[htbp]
		\centering
		\caption{\textit{Mean of the posterior probabilities using integral priors, intrinsic priors, and mean of the probabilities of the models using Akaike weights.}}
		\begin{tabular}{ccccccccccccccccccc}
			\hline
			&       & \multicolumn{17}{c}{True model: $M_1$ with $\theta_1=0.5$} \\
			\cline{3-19}          &       & \multicolumn{5}{c}{Integral priors}   &       & \multicolumn{5}{c}{Intrinsic priors}  &       & \multicolumn{5}{c}{Akaike weights} \\
			\cline{3-7}\cline{9-13}\cline{15-19}    $n$   &       & $M_1$ & $M_2$ & $M_3$ & $M_4$ & $M_5$ &       & $M_1$ & $M_2$ & $M_3$ & $M_4$ & $M_5$ &       & $M_1$ & $M_2$ & $M_3$ & $M_4$ & $M_5$ \\
			15    &       & 0.25  & 0.15  & 0.19  & 0.20  & 0.21  &       & 0.23  & 0.17  & 0.19  & 0.20  & 0.21  &       & 0.24  & 0.15  & 0.19  & 0.21  & 0.21 \\
			50    &       & 0.29  & 0.10  & 0.18  & 0.21  & 0.23  &       & 0.29  & 0.10  & 0.17  & 0.21  & 0.23  &       & 0.31  & 0.09  & 0.17  & 0.21  & 0.23 \\
			100   &       & 0.35  & 0.06  & 0.15  & 0.21  & 0.24  &       & 0.37  & 0.04  & 0.14  & 0.20  & 0.24  &       & 0.38  & 0.04  & 0.14  & 0.20  & 0.24 \\
			300   &       & 0.59  & 0.00  & 0.05  & 0.14  & 0.22  &       & 0.59  & 0.00  & 0.06  & 0.14  & 0.21  &       & 0.60  & 0.00  & 0.05  & 0.13  & 0.21 \\
			500   &       & 0.66  & 0.00  & 0.03  & 0.11  & 0.19  &       & 0.67  & 0.00  & 0.03  & 0.11  & 0.19  &       & 0.68  & 0.00  & 0.03  & 0.10  & 0.19 \\
			&       &       &       &       &       &       &       &       &       &       &       &       &       &       &       &       &       &  \\
			&       & \multicolumn{17}{c}{True model: $M_1$ with $\theta_1=3$} \\
			\cline{3-19}          &       & \multicolumn{5}{c}{Integral priors}   &       & \multicolumn{5}{c}{Intrinsic priors}  &       & \multicolumn{5}{c}{Akaike weights} \\
			\cline{3-7}\cline{9-13}\cline{15-19}    $n$   &       & $M_1$ & $M_2$ & $M_3$ & $M_4$ & $M_5$ &       & $M_1$ & $M_2$ & $M_3$ & $M_4$ & $M_5$ &       & $M_1$ & $M_2$ & $M_3$ & $M_4$ & $M_5$ \\
			15    &       & 0.52  & 0.02  & 0.08  & 0.16  & 0.22  &       & 0.51  & 0.02  & 0.09  & 0.16  & 0.21  &       & 0.61  & 0.01  & 0.06  & 0.13  & 0.19 \\
			50    &       & 0.87  & 0.00  & 0.00  & 0.03  & 0.09  &       & 0.85  & 0.00  & 0.01  & 0.04  & 0.10  &       & 0.85  & 0.00  & 0.00  & 0.04  & 0.11 \\
			100   &       & 0.93  & 0.00  & 0.00  & 0.01  & 0.05  &       & 0.97  & 0.00  & 0.00  & 0.01  & 0.03  &       & 0.95  & 0.00  & 0.00  & 0.01  & 0.05 \\
			300   &       & 1.00  & 0.00  & 0.00  & 0.00  & 0.00  &       & 1.00  & 0.00  & 0.00  & 0.00  & 0.00  &       & 1.00  & 0.00  & 0.00  & 0.00  & 0.00 \\
			&       &       &       &       &       &       &       &       &       &       &       &       &       &       &       &       &       &  \\
			&       & \multicolumn{17}{c}{True model: $M_2$ with $\theta_2=0.2$} \\
			\cline{3-19}          &       & \multicolumn{5}{c}{Integral priors}   &       & \multicolumn{5}{c}{Intrinsic priors}  &       & \multicolumn{5}{c}{Akaike weights} \\
			\cline{3-7}\cline{9-13}\cline{15-19}    $n$   &       & $M_1$ & $M_2$ & $M_3$ & $M_4$ & $M_5$ &       & $M_1$ & $M_2$ & $M_3$ & $M_4$ & $M_5$ &       & $M_1$ & $M_2$ & $M_3$ & $M_4$ & $M_5$ \\
			15    &       & 0.01  & 0.53  & 0.24  & 0.14  & 0.09  &       & 0.01  & 0.52  & 0.24  & 0.14  & 0.10  &       & 0.01  & 0.47  & 0.25  & 0.16  & 0.11 \\
			50    &       & 0.00  & 0.84  & 0.13  & 0.02  & 0.01  &       & 0.00  & 0.78  & 0.18  & 0.03  & 0.01  &       & 0.00  & 0.76  & 0.20  & 0.04  & 0.01 \\
			100   &       & 0.00  & 0.94  & 0.06  & 0.00  & 0.00  &       & 0.00  & 0.94  & 0.06  & 0.00  & 0.00  &       & 0.00  & 0.93  & 0.07  & 0.00  & 0.00 \\
			&       &       &       &       &       &       &       &       &       &       &       &       &       &       &       &       &       &  \\
			&       & \multicolumn{17}{c}{True model: $M_2$ with $\theta_2=0.5$} \\
			\cline{3-19}          &       & \multicolumn{5}{c}{Integral priors}   &       & \multicolumn{5}{c}{Intrinsic priors}  &       & \multicolumn{5}{c}{Akaike weights} \\
			\cline{3-7}\cline{9-13}\cline{15-19}    $n$   &       & $M_1$ & $M_2$ & $M_3$ & $M_4$ & $M_5$ &       & $M_1$ & $M_2$ & $M_3$ & $M_4$ & $M_5$ &       & $M_1$ & $M_2$ & $M_3$ & $M_4$ & $M_5$ \\
			15    &       & 0.13  & 0.29  & 0.22  & 0.19  & 0.18  &       & 0.11  & 0.34  & 0.21  & 0.18  & 0.16  &       & 0.12  & 0.31  & 0.21  & 0.18  & 0.17 \\
			50    &       & 0.04  & 0.46  & 0.23  & 0.15  & 0.12  &       & 0.04  & 0.51  & 0.22  & 0.13  & 0.10  &       & 0.04  & 0.49  & 0.23  & 0.14  & 0.10 \\
			100   &       & 0.01  & 0.66  & 0.20   & 0.08  & 0.05  &       & 0.00  & 0.64  & 0.22  & 0.09  & 0.05  &       & 0.00  & 0.65  & 0.20  & 0.09  & 0.05 \\
			300   &       & 0.00  & 0.89  & 0.10  & 0.01  & 0.00  &       & 0.00  & 0.89  & 0.10  & 0.01  & 0.00  &       & 0.00  & 0.86  & 0.13  & 0.01  & 0.00 \\
			&       &       &       &       &       &       &       &       &       &       &       &       &       &       &       &       &       &  \\
			&       & \multicolumn{17}{c}{True model: $M_3$ with $\theta_2=0.2$} \\
			\cline{3-19}          &       & \multicolumn{5}{c}{Integral priors}   &       & \multicolumn{5}{c}{Intrinsic priors}  &       & \multicolumn{5}{c}{Akaike weights} \\
			\cline{3-7}\cline{9-13}\cline{15-19}    $n$   &       & $M_1$ & $M_2$ & $M_3$ & $M_4$ & $M_5$ &       & $M_1$ & $M_2$ & $M_3$ & $M_4$ & $M_5$ &       & $M_1$ & $M_2$ & $M_3$ & $M_4$ & $M_5$ \\
			15    &       & 0.00  & 0.17  & 0.33  & 0.28  & 0.22  &       & 0.01  & 0.20  & 0.31  & 0.27  & 0.22  &       & 0.01  & 0.16  & 0.30  & 0.28  & 0.24 \\
			50    &       & 0.00  & 0.12  & 0.55  & 0.24  & 0.08  &       & 0.00  & 0.09  & 0.60  & 0.23  & 0.08  &       & 0.00  & 0.07  & 0.59  & 0.25  & 0.09 \\
			100   &       & 0.00  & 0.03  & 0.74  & 0.21  & 0.03  &       & 0.00  & 0.02  & 0.81  & 0.16  & 0.01  &       & 0.00  & 0.01  & 0.80  & 0.17  & 0.02 \\
			&       &       &       &       &       &       &       &       &       &       &       &       &       &       &       &       &       &  \\
			&       & \multicolumn{17}{c}{True model: $M_3$ with $\theta_2=0.5$} \\
			\cline{3-19}          &       & \multicolumn{5}{c}{Integral priors}   &       & \multicolumn{5}{c}{Intrinsic priors}  &       & \multicolumn{5}{c}{Akaike weights} \\
			\cline{3-7}\cline{9-13}\cline{15-19}    $n$   &       & $M_1$ & $M_2$ & $M_3$ & $M_4$ & $M_5$ &       & $M_1$ & $M_2$ & $M_3$ & $M_4$ & $M_5$ &       & $M_1$ & $M_2$ & $M_3$ & $M_4$ & $M_5$ \\
			15    &       & 0.16  & 0.18  & 0.22  & 0.22  & 0.21  &       & 0.12  & 0.22  & 0.23  & 0.22  & 0.20  &       & 0.15  & 0.19  & 0.22  & 0.22  & 0.22 \\
			50    &       & 0.03  & 0.13  & 0.32  & 0.29  & 0.23  &       & 0.05  & 0.18  & 0.30  & 0.26  & 0.21  &       & 0.04  & 0.16  & 0.32  & 0.27  & 0.21 \\
			100   &       & 0.00  & 0.14  & 0.43  & 0.27  & 0.16  &       & 0.00  & 0.15  & 0.46  & 0.25  & 0.14  &       & 0.00  & 0.13  & 0.43  & 0.28  & 0.17 \\
			300   &       & 0.00  & 0.01  & 0.70  & 0.24  & 0.05  &       & 0.00  & 0.01  & 0.70  & 0.24  & 0.05  &       & 0.00  & 0.02  & 0.70  & 0.23  & 0.05 \\
			\hline
		\end{tabular}%
		\label{tab:addlabel}%
	\end{table}%
}

\section{Integral posteriors and evidence approximation}
In case copycat models with compact parametric spaces have been added, a simple Monte Carlo estimator of the Bayes factor for integral priors is based on the Markov chains $(\theta_i^t)$ associated with integral priors $\pi_i(\theta_i)$, $i=1,\dots,q$. For example, we can estimate the Bayes factor $B_{21}(x)$ by
$\sum_{t=1}^T f_2(x\mid\theta_2^t)/\sum_{t=1}^T f_1(x\mid\theta_1^t)$.
However, it is well known that this is not the
optimal strategy when the integral posterior $\pi_i(\theta_i\mid x)\propto\pi_i(\theta_i)f_i(x\mid\theta_i)$ is concentrated relative to the integral prior $\pi_i(\theta_i)$. In such situations general methods based on the integral posterior could produce estimates of the evidence, and therefore of the Bayes factor, in a more efficient way. Besides, the simulation from the integral posterior entails some additional difficulties because the integral prior is not known analytically. To overcome this issue we resort to a completion of the integral posterior as follows.

The simulation of $\theta_{i}^{t+1}$ given $\theta_{i}^{t}$ is carried out in several steps. The last two steps correspond to the simulation of a training sample $z_i^{t+1}$, and the simulation $\theta_{i}^{t+1}$ from $\pi_i^N(\theta_i\mid z_i^{t+1})$. 
%In this way the process $(z_i^t)$ arises. 
Besides, since the 2-step transition density $Q_i^2(\theta_i\mid\theta_i^{\prime})$ does not depend on $\theta_i^{\prime}$, it follows that given any initial point $\theta_i^0$, the subsequences $\{\theta_i^2$, $\theta_i^4$, $\theta_i^6$, $\dots\}$ and $\{\theta_i^3$, $\theta_i^5$, $\theta_i^7$, $\dots\}$ are iid from $\pi_i(\theta_i)$. Note that $\theta_i^t$ and $\theta_i^{t+1}$ are independent if any compact space is involved along the transition $\theta_i^t\rightarrow\theta_i^{t+1}$. In general, we consider the subsequence $(\check{\theta}_i^t)$ of the Markov chain $(\theta_i^t)$ such that $\check{\theta}_i^t$ and  $\check{\theta}_i^{t+1}$ are independent, and the corresponding subsequence $(\check{z}_i^t)$ of $(z_i^t)$. This sequence is the key ingredient of the completion method. If $p_i(z_i)$ is the stationary distribution of $(z_i^t)$, then $(\check{z}_i^t)$ are iid from $p_i(z)$, and
\[
\pi_i(\theta_i)=\int \pi_i^N(\theta_i\mid z_i)p_i(z_i)\mathrm{d}z_i.
\]
Therefore
\[
\pi_i(\theta_i,z_i\mid x)=\frac{f_i(x\mid \theta_i)\pi_i^N(\theta_i\mid z_i)p_i(z_i)}{m_i(x)}
\]
is a completion of $\pi_i(\theta_i\mid x)$, and $\pi_i(\theta_i\mid z_i,x)=\pi_i^N(\theta_i\mid z_i,x)$. Then the Metropolis-Hastings algorithm with target $\pi_i(\theta_i,z_i\mid x)$ and proposal of the form
\[
q_i(\theta_i^{\prime},z_i^{\prime}\mid \theta_i,z_i,x)=p_i(z_i^{\prime})\rho_i(\theta_i^{\prime}\mid \theta_i,z_i^{\prime},x),
\]
can be implemented provided
\begin{equation}
	\frac{f_i(x\mid\theta_i^{\prime})\pi_i^N(\theta_i^{\prime}\mid z_i^{\prime})\rho_i(\theta_i\mid \theta_i^{\prime},z_i,x)}{f_i(x\mid\theta_i)\pi_i^N(\theta_i\mid z_i)\rho_i(\theta_i^{\prime}\mid \theta_i,z_i^{\prime},x)}.\label{acceptance}
\end{equation}
can be computed. Note that if $\rho_i(\theta_i^{\prime}\mid \theta_i,z_i^{\prime},x)$ does not depend on $z_i^{\prime}$, then this Metropolis-Hasting algorithm becomes the pseudo-marginal Metropolis-Hastings algorithm of \cite{AndrieuRoberts2009} for the target $\pi_i(\theta_i\mid x)$, where the unbiased estimator of the target is $\pi_i(\theta_i,z_i\mid x)/p_i(z_i)$. If we choose
\[
\rho_i(\theta_i^{\prime}\mid\theta_i,z_i^{\prime},x)=\pi_i^N(\theta_i^{\prime}\mid z_i^{\prime},x)=\frac{f_i(x\mid\theta_i^{\prime})\pi_i^N(\theta_i^{\prime}\mid z_i^{\prime})}{m_i^N(z_i^{\prime},x)/m_i^N(z_i^{\prime})},
\]
then (\ref{acceptance}) is given by
\[
\frac{m_i^N(z_i^{\prime},x)}{m_i^N(z_i^{\prime})}\frac{m_i^N(z_i)}{m_i^N(z_i,x)}.
\]
The aforementioned Metropolis-Hastings algorithm produces a Markov chain $(\tilde{\theta}_i^t,\tilde{z}_i^t)$ through the following transition. Given $\tilde{\theta}_i^t$ and $\tilde{z}_i^t$:

\begin{enumerate}
	\item Take $z_i^{\prime}=\check{z}_i^{t+1}$ and generate $\theta_i^{\prime}\sim \rho_i(\theta_i^{\prime}\mid \tilde{\theta}_i^t,\check{z}_i^{t+1},x)$
	
	\item Take $\tilde{\theta}_i^{t+1}=\theta_i^{\prime}$ and $\tilde{z}_i^{t+1}=z_i^{\prime}$ with probability
	\[
	1 \wedge
	\frac{f_i(x\mid\theta_i^{\prime})\pi_i^N(\theta_i^{\prime}\mid z_i^{\prime})\rho_i(\tilde{\theta}_i^t\mid \theta_i^{\prime},\tilde{z}_i^t,x)}{f_i(x\mid\tilde{\theta}_i^t)\pi_i^N(\tilde{\theta}_i^t\mid \tilde{z}_i^t)\rho_i(\theta_i^{\prime}\mid \tilde{\theta}_i^t,z_i^{\prime},x)},
	\]
	otherwise take $\tilde{\theta}_i^{t+1}=\tilde{\theta}_i^{t}$ and $\tilde{z}_i^{t+1}=\tilde{z}_i^{t}$.
\end{enumerate}
Then the integral posterior and the integral prior can be estimated by
\[
\hat{\pi}_i(\theta_i\mid x)=\frac{1}{\tilde{T}}\sum_{t=1}^{\tilde{T}}\pi_i^N(\theta_i\mid \tilde{z}_i^t,x),
\]
and
\[
\hat{\pi}_i(\theta_i)=\frac{1}{T}\sum_{t=1}^T\pi_i^N(\theta_i\mid z_i^t),
\]
respectively.

This allows us to use Chib’s (\citeyear{Chib1995}) approach to estimate the evidence by
\begin{equation}
	\frac{\hat{\pi}_i(\theta_i^*)f_i(x\mid\theta_i^*)}{\hat{\pi}_i(\theta_i^*\mid x)},\label{chib}
\end{equation}
where $\theta_i^*$ is any plug-in value for $\theta_i$. 

\begin{figure}
	\centering
	\includegraphics[width=0.7\linewidth]{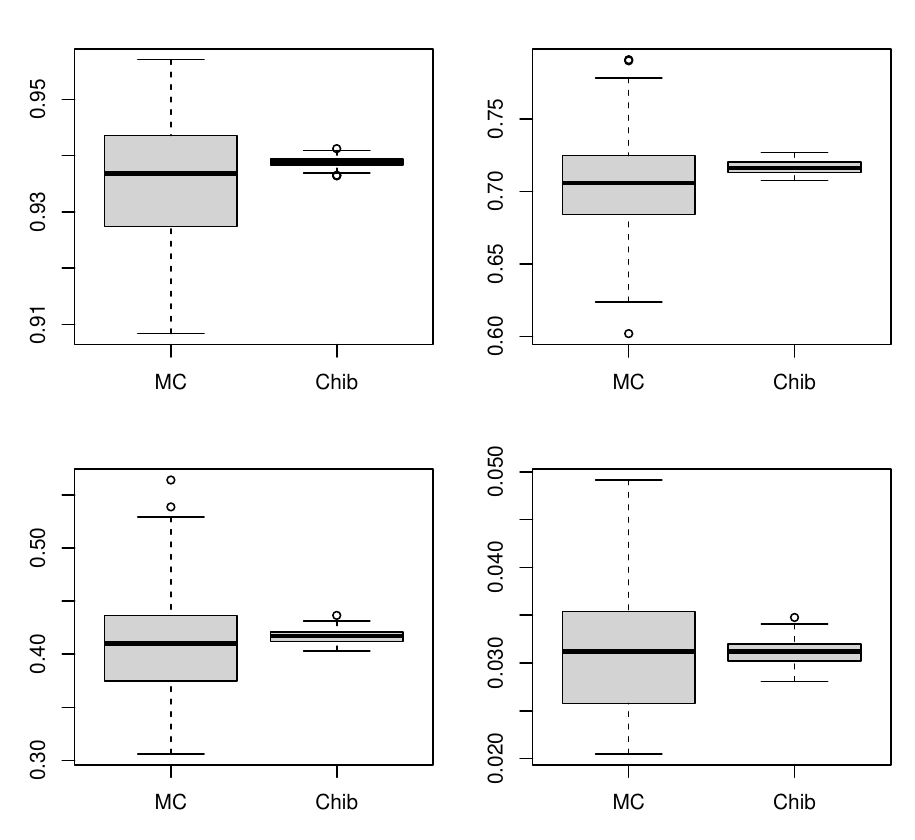}
	\caption{\textit{Comparison of the estimates of the posterior probability of model $M_1$ (last row in Table \ref{tablaANOVA}) using the estimator MC (\ref{MCestimator}), and the Chib's method with the proposed Metropolis-Hastings algorithm to simulate from the integral posterior}.}
	\label{chibfig}
\end{figure}

We have applied this procedure to replicate the last row in Table \ref{tablaANOVA} using the Monte Carlo estimator (\ref{MCestimator}) and the Chib's method with the simulations from the integral posteriors with proposal $\rho_i(\theta_i^{\prime}\mid\theta_i,z_i^{\prime},x)=\pi_i^N(\theta_i^{\prime}\mid z_i^{\prime},x)$ in the Metropolis-Hastings algorithm. The number of simulations was set to $T=10,000$ to approximate the integral priors and $\tilde{T} = 5000$ to approximate the integral posteriors. Using
these approximations the evidence was estimated according to equation \ref{chib}, and the estimation was repeated 100 times. The resulting boxplots in Figure \ref{chibfig} illustrate 
the corresponding behavior and the superiority of Chib's method.

\section{Conclusions and oncoming research}
We advocated in this paper a principled generalization of the methodology of integral priors  \citep{Cano2008} when two or more models are to be compared. Our approach proves to be quite automated and generic, both in the nested and in the non-nested cases, with the only requirement being the existence of training samples of finite size for all models under comparison. Although we have carried out some comparisons with more approximate solutions like intrinsic and fractional pseudo Bayes factors \citep{Robert2001}, further experiments could be conducted in that spirit. We recognize that the application of integral priors for complex models could involve certain degree of difficulty with respect to the concept of imaginary training samples, similar to the ones encountered with other automatic methods like expected posterior priors or intrinsic priors. On the other hand, usually integral priors have not an explicit form but must be obtained by simulation, which entails some computational limitations.

This methodology is an original proposal for the problem of choosing priors for model selection in the general case, rather than suggesting priors for specific cases. In this sense, our methodology falls within the class of approaches that produce default priors. Other methods, such as intrinsic priors, have been shown to work well for nested models, whereas for general non-nested models the problem has to be studied case-by-case. Our methodology is different in that it addresses all models simultaneously without privileging any one over another. However, we acknowledge that there is still much work to be done in the development of this theory, for example, improving computation time when the number of models is large, as well as the estimation of the evidence.

The contribution of our work, in relation to the prior developments by \cite{Cano2013} and \cite{Cano2018}, lies in having established conditions under which the integral priors are unique and proper prior distributions in such a way that the proposed methodology enables the application of the Metropolis-Hastings algorithm to simulate from the posterior distribution, thereby facilitating the use of general methods such as the harmonic mean estimator or Chib’s method \citep{Chib1995} to approximate the \textit{evidence}, and consequently, the Bayes factor.

The examples we considered so far allow to approximate the Bayes factor with the Monte Carlo estimator, but extensions to more general settings are directly available when using evidence approximations \citep{FrielWyse2011}, like bridge sampling, inverse logistic regression or even harmonic means since some models \citep{MarinRobert2011} are associated with compact parameter spaces.

Occasionally researchers can express their knowledge about the parameters
by establishing a range of possible values. In such cases, it would be possible to work
with proper priors. If any of the models considered is equipped with such a proper prior, then it is not necessary to use artificial models with compact parametric spaces because the original model does not require imaginary training samples, and the prior would be the integral prior. Furthermore, we could use these ranges of values to construct the artificial models, in which case the integral priors could differ.
\\

\textbf{Acknowledgements}
The research of D Salmer\'on was supported by the Fundación SéNeCa-Agencia de Ciencia y Tecnología de la Región de Murcia Program for Excellence in Scientific Research (project 20862/PI/18). CP Robert is partially funded by the European Union (ERC-2022-SYG-OCEAN-101071601) and by a PR[AI]RIE-PSAI chair from the Agence Nationale de la Recherche (ANR-23-IACL-0008).
%Prairie chair from the Agence Nationale de la Recherche (ANR-19-P3IA-0001).

\vspace{1cm}

\textbf{Juan Antonio Cano} (1956-2018) was Professor in the Department of Statistics and Operations Research at the University of Murcia. Dr. Cano was a dear mentor and friend who contributed substantially to the theory of integral priors and was instrumental to the developments of this article.

\newpage	
\section*{Supplementary material}
\subsection*{Data for Example 1} The sample for the example 1 was $x=\{$1.094, 0.105, 1.315, 0.655, 0.860, 1.630, 0.275, 0.634, 0.377, 0.667, 1.327, 0.002, 0.535, 0.230, 0.795$\}$.

\subsection*{Data for Section 4} $\{$-6.08601161, -0.38629099, -0.94318682, -1.10768258, 1.62898983, -4.53073160, -0.65574138, -0.58995846, 3.85144493, -3.94493171, 1.32760319, -0.84059533, -2.38323565, -0.08982564, 0.84510522$\}$.

\end{document}